\documentclass[%
  aps,%
  prl,%
  twocolumn,%
  floatfix, %
  superscriptaddress,%
  longbibliography,% keeps full titles/author lists in thebibliography
  nofootinbib,% keep \footnote{}s at the bottom of the page (not in the bibliography)
]{revtex4-2}
\usepackage{xr-hyper}
\makeatletter
\def\doauthor#1#2#3{%
  \ignorespaces#1\unskip
  \begingroup
   #3% prints affiliation superscripts
  \@if@empty{#2}{\endgroup{}{}}{\endgroup{\comma@space}{}\frontmatter@footnote{#2}}%
  \@listcomma
  \space \@listand
}%
\makeatother

\usepackage{amsmath,amssymb}
\usepackage{bbm} % for blackboard-bold digits like \mathbbm{1}
\usepackage{graphicx}
\usepackage{dblfloatfix} % better bottom placement for two-column floats (figure*/table*)
\usepackage{capt-of} % provides \captionof{figure}{...} outside floating environments
\usepackage[caption=false]{subfig}
\usepackage{algpseudocode}

\newcounter{algorithm}
\renewcommand{\thealgorithm}{\arabic{algorithm}}

\usepackage{booktabs}
\usepackage{xcolor}
\usepackage{bm}
\usepackage[colorlinks=true,linkcolor=blue,citecolor=blue,urlcolor=blue]{hyperref}

\makeatletter
\let\revtex@section\section
\renewcommand\section{\@startsection{section}{1}{\z@}%
  {0.6ex \@plus .2ex \@minus .2ex}%  % space before
  {-1em}%                              % run-in heading
  {\itshape}}%
\makeatother

\AtBeginDocument{%
  \setlength{\abovedisplayskip}{2pt plus 1pt minus 1pt}%
  \setlength{\belowdisplayskip}{2pt plus 1pt minus 1pt}%
  \setlength{\abovedisplayshortskip}{0pt plus 1pt minus 1pt}%
  \setlength{\belowdisplayshortskip}{0pt plus 1pt minus 1pt}%
}

\usepackage{graphicx}
\usepackage{booktabs}
\usepackage{multirow}

\usepackage{xcolor}
\usepackage{bm}
\usepackage{hyperref}

\definecolor{todo}{HTML}{cc0000}

\newcommand{\bphi}{\bm{\phi}}
\newcommand{\beps}{\bm{\epsilon}}
\newcommand{\bw}{\mathbf{w}}

\newcommand{\s}{\mathbf{s}}
\newcommand{\bx}{\mathbf{x}}

\newcommand{\bmu}{\bm{\mu}}
\newcommand{\bth}{\bm{\theta}}

\newcommand{\Id}{\mathbf{I}}
\newcommand{\zero}{\mathbf{0}}

\newcommand{\E}{\mathbb{E}}

\makeatletter

\renewcommand{\thealgorithm}{\@arabic\c@algorithm}

\newenvironment{algorithm}[1][tbp]{%
  \begin{figure}[#1]%
  \renewcommand{\caption}[1]{%
    \refstepcounter{algorithm}%
    \par\noindent\textbf{Algorithm \thealgorithm:}\ ##1\par\smallskip}%
  \hrule height.9pt\kern2pt}{%
  \kern2pt\hrule\relax\end{figure}}
\makeatother

\begin{document}

\title{Data Free Training of Diffusion Models for Lattice Field Theories}

\author{Dennis Grinwald}
\email[Correspondence: ]{dennis.grinwald@tu-berlin.de}
\affiliation{Berlin Institute for the Foundations of Learning and Data (BIFOLD), Germany}
\affiliation{Technische Universit\"{a}t Berlin, Germany}

\author{Ankur Singha}
\email[Correspondence: ]{a.singha@tu-berlin.de}
\affiliation{Berlin Institute for the Foundations of Learning and Data (BIFOLD), Germany}
\affiliation{Technische Universit\"{a}t Berlin, Germany}

\author{Khaled Kahouli}
%\email[Correspondence: ]{khaled.kahloui}
\affiliation{Berlin Institute for the Foundations of Learning and Data (BIFOLD), Germany}
\affiliation{Technische Universit\"{a}t Berlin, Germany}

\author{Karl Jansen}
\affiliation{Deutsches Elektronen-Synchrotron (DESY), Platanenallee 6,
15738 Zeuthen, Germany}
\affiliation{Computation-Based Science and Technology Research Center, The Cyprus Institute, Nicosia, Cyprus}

\author{Shinichi Nakajima}
\email[Correspondence: ]{nakajima@tu-berlin.de}
\affiliation{Berlin Institute for the Foundations of Learning and Data (BIFOLD), Germany}
\affiliation{Technische Universit\"{a}t Berlin, Germany}
\affiliation{RIKEN Center for Advanced Intelligence Project (AIP), Japan}

\date{\today}

\begin{abstract}
Sampling equilibrium field configurations remains a major computational bottleneck in lattice field theory, motivating the development of generative samplers that can produce independent configurations. Diffusion models are promising for this purpose, but their standard denoising score matching objective requires samples from the target distribution, seemingly reintroducing the original sampling problem. Here, we show that diffusion models can instead be trained from their own generated samples using the target score identity, provided access to the unnormalized probability density.  Importance-weighted formulations of this identity yield a consistent target throughout training, while unweighted variants define a self-consistent fixed-point training scheme. We benchmark the different approaches in two-dimensional \(\phi^4\) theory and identify effective choices that enable efficient training and sampling across the symmetric, critical, and broken phases. Ultimately, our data-free training approach achieves comparable performance to training on HMC reference data. 
\end{abstract}

\maketitle

\section{Introduction.}

Lattice field theory (LFT) is a cornerstone of the study of non-perturbative quantum field theories~\cite{WilsonKogut1974,creutz1983quarks,CREUTZ1983201,kronfeld2012twenty}. In lattice quantum chromodynamics in particular, first-principles simulations deliver quantities of direct phenomenological relevance such as quark masses, meson decay constants, and the strong coupling constant, all at controlled precision~\cite{flag2024,komijani2020strong}. These quantities are obtained from expectation values of observables, evaluated by sampling field configurations according to the Boltzmann distribution defined by the lattice action.
The standard framework for this task is Markov chain Monte Carlo (MCMC), most commonly Hybrid Monte Carlo (HMC)~\cite{duane1987hybrid}.
However, as one approaches the critical region or continuum limit of the theory, MCMC suffers from \emph{critical slowing down~\cite{Wolff:1989wq,wolff2004monte,Schaefer:2010hu}} due to the long range correlation where the integrated autocorrelation time and the statistical error on some important  observable grows at fixed chain length. Producing a fixed number of effectively independent configurations therefore becomes rapidly more expensive precisely in the regimes of physical interest.

Generative samplers offer a promising route to mitigating critical slowing down by directly producing statistically independent field configurations ~\cite{kobyzev2020normalizing,ho2020ddpm,nicoli_pre,PhysRevD.100.034515,Wang:2023exq,singha2025multilevel}.
% This bottleneck can be removed by \emph{independent} generative samplers~\cite{kobyzev2020normalizing,ho2020ddpm,PhysRevD.100.034515,Wang:2024xyz,singha2025multilevel} that draw statistically uncorrelated configurations directly, such as normalizing flows~\cite{PhysRevD.100.034515,Kanwar:2020xzo,nicoli_prl,Caselle:2022esc,Singha:2023cql,10.21468/SciPostPhysCore.5.4.052,Nicoli:2023qsl,Faraz:2023xdi,bauer2025super,singha2026multiscale,Abbott:2022zsh,gerdes2023learning,Singha:2023xxq,Finkenrath:2024pR,Kreit:2026eng,Schuh:2026dvp}.
Substantial progress has been made with normalizing flows, from sampling scalar field theories ~\cite{PhysRevD.100.034515,nicoli_prl,Caselle:2022esc,caselle2024sampling,Singha:2023cql,10.21468/SciPostPhysCore.5.4.052,Nicoli:2023qsl,10.21468/SciPostPhys.16.5.132,Faraz:2023xdi,bauer2025super,singha2026multiscale,gerdes2023learning} to constructing samplers for lattice gauge theories~\cite{Kanwar:2020xzo,Abbott:2022zsh,Singha:2023xxq,Abbott:2023thq,Bacchio:2022vje,Gerdes:2024rjk,Finkenrath:2024pR,singha2026multilevel,Bulgarelli:2024cqc} and developing approaches to address topological freezing~\cite{Kanwar:2020xzo,Singha:2023xxq,bonanno2026scalable,Singha:2026aac}. Despite these advances, the computational cost and architectural requirements of normalizing flows remain important limitations. Coupling-based flows impose architectural constraints to retain tractable density evaluation, while continuous normalizing flows offer greater architectural flexibility at the cost of expensive numerical integration and divergence evaluation during training.
% and diffusion (score-based) models~\cite{Wang:2024xyz,Zhu:2025pmw,aart,Kanwar:2025wuc,vega2025group,Kanaujia:2025jru,Alharazin:2026lcb}. Both families have a cost of their own. Coupling flows are restricted in architecture, thus sacrificing model expressivity, while continuous flow~\cite{Bacchio:2022vje,caselle2024sampling,Gerdes:2024rjk} requires computation of probability densities at every training step, which is expensive and scales poorly with lattice volume~\cite{Abbott:2022zsh}. 

% Score-based diffusion models avoid density computation during training but must be trained on a large set of pre-generated configurations~\cite{Wang:2024xyz,vega2025equivariant}, reintroducing the very MCMC cost one is trying to avoid.

Diffusion models provide an alternative to these flow-based approaches. Their success in machine learning applications ~\cite{ho2020ddpm,song2021scorebased} has motivated growing interest in lattice field theory, with applications to scalar field theories~\cite{Wang:2023exq,vega2025equivariant,tan2026diffusion} and lattice gauge theories~\cite{Zhu:2025pmw,Kanwar:2025wuc,aart,komijani2026diffusion,Alharazin:2026lcb,vega2026sampling}. 
A central obstacle to the practical use of diffusion models in lattice field theory is their reliance on representative training data or computationally expensive annealing procedures. Generating training ensembles with MCMC can reintroduce the original sampling bottleneck, while annealing-based approaches incur additional costs through sampling and correction steps across intermediate distributions. These limitations motivate a simple and computationally efficient data-free training approach that learns directly from the lattice action.

% A recent line of work \TODO{ADD QC works} removes both costs at once by training diffusion \emph{samplers} that need neither computation of probability densities nor training data, using only pointwise evaluations of the unnormalized density $e^{-S}$ and its gradient. 
% Iterative importance weighted variants of target score matching (TSM)~\cite{de2024target} such as Iterated Denoising Energy Matching (iDEM)~\cite{akhoundsadegh2024idem} are the prototypical examples: they first fill a replay buffer with the current model's own draws and second regress the score against a suitable target. That estimator relies on self-normalized importance weights whose variance grows sharply with dimension. On $\phi^4$ at larger lattice volumes this makes training unstable and, in practice, infeasible. 

% Alternative data-free samplers trade this instability for a heavier training procedure: the Stochastic Path Sampler (SPS)~\cite{chen2026sps} learns the noise schedule as well as forward and backward drifts by minimizing a path-space, trajectory-level objective, and so must simulate and store full trajectories during training. 
% % \TODO{ADD DATA FREE DIFFUSION PAPER HERE}. 
% Annealing based approaches such as Progressive Inference-Time Annealing (PITA)~\cite{akhoundsadegh2025pita}, sidesteps the hard low-temperature target by sampling at high temperature where the modes are easy to reach and annealing the trained model down to the target temperature via Feynman-Kac correction.
Recent work in the machine learning community has introduced data-free approaches to training diffusion samplers, including methods based on reverse Kullback–Leibler divergence over path measures \cite{zhang2021path, berner2024an, vargas2023denoising}, computationally expensive MCMC guided by the denoising score identity (DSI) \cite{huang2023reverse,grenioux2024stochastic}, and target score matching (TSM)~\cite{de2024target}, which relies on the target density and its gradient. However, their efficient application to lattice field theory remains challenging. For instance, importance-weighted methods, such as iDEM~\cite{akhoundsadegh2024idem} are sensitive to poor overlap between proposal and target distributions, which can become increasingly severe as the lattice volume grows. Other approaches incur additional computational costs through trajectory-based training, as in SPS~\cite{chen2026sps} or sampling and correction across intermediate distributions, as in PITA~\cite{akhoundsadegh2025pita}. 

% In this work, we build on the iterative training procedure introduced by iDEM~\cite{akhoundsadegh2024idem} where the model supplies its own training configurations through a replay buffer, which gets frequently updated. Unlike iDEM, we \emph{do not} use importance weights. Instead, we regress against the data samples directly. Moreover, we regress against the control-variate score-matching target~\cite{kahouli2025cvsm} paired with a noise schedule normalized to total-variance. The target optimally blends the two score identities so that the variance of the score target is minimized as shown in Fig.~\ref{fig:estimator-variance}. The variance-minimizing blend coefficient is available in closed form. These two ingredients stabilize the training procedure and make data-free score-based sampling feasible in larger lattices where the importance-weighted objective breaks down.
In this work, we establish TSM as a practical framework for data-free diffusion sampling in lattice field theory. TSM relates the diffusion score to the target-density gradient, allowing the lattice action to supply the learning signal. The model generates and periodically refreshes its own training configurations, as in iDEM~\cite{akhoundsadegh2024idem} but learns through an unweighted, self-consistent objective that avoids importance-weight fluctuations during training, To stabilize this learning process, we combine a control-variate score target~\cite{kahouli2025cvsm}, which balances action-gradient and denoising information through a closed-form variance-minimizing coefficient, with a noise schedule normalized to fixed total variance. The resulting training procedure requires neither precomputed target ensembles nor a model-density evaluation.

% We demonstrate our method on two-dimensional $\phi^4$ theory for lattice sizes $L\in\{4,8,16\}$ in the symmetric, critical, and broken phases. 
% Because the sampler admits likelihood computation through its probability-flow ODE, every configuration carries an importance weight to the exact target, yielding self-consistent free-energy estimates and reweighted observables. 
We demonstrate the approach in two-dimensional scalar field theory across the symmetric, critical, and broken phases, obtaining asymptotically unbiased estimates of physical observables through importance reweighting. At criticality, the sampler achieves efficiencies comparable to the corresponding diffusion models trained on HMC ensembles, demonstrating that reference data are not required to achieve this sampling performance. The unweighted training scheme enables stable learning and efficient sampling at lattice volumes where importance-weighted training becomes unstable. These results establish a practical route to learning equilibrium lattice distributions directly from the action, retaining accurate physical predictions while avoiding the costs of generating MCMC training ensembles and performing additional annealing procedures.
\section{Method.}
% Our lattice field theories are defined on a hypercube lattice $\Lambda_L=\{ 1,...,L \}^D$ with continuous degrees of freedom $\phi_{\bx}\in\mathbb{R}$ at each site $\bx\in\Lambda_L$. 
We consider a real scalar field on a \(D\)-dimensional hypercubic lattice \(\Lambda_L=\{1,\ldots,L\}^{D}\), with a field variable \(\phi_{\mathbf{x}}\in\mathbb{R}\) at each site \(\mathbf{x}\in\Lambda_L\).
A field configuration is the collection $\bphi=\{\phi_{\bx}\}_{\bx\in\Lambda_L}\in\mathbb{R}^{|\Lambda_L|}$. The action over field configurations is defined as:
\begin{align}
    S[\bphi] = \sum_{\bx}\big[ -2\kappa\sum_\mu\phi_{\bx}\phi_{\bx+\hat{\bmu}} + (1-2\lambda)\phi_{\bx}^2 + \lambda\phi_{\bx}^4 \big].
\end{align}
% We wish to sample from the Boltzmann distribution over configurations
% We aim to sample field configurations from the Boltzmann distribution:
Our goal is to sample field configurations from the Boltzmann distribution:
\begin{align}
    p(\bphi)=\frac{1}{Z}e^{-S[\bphi]}, \mbox{ where } Z=\int \mathcal{D}\bphi\, e^{-S[\bphi]},  \mathcal{D}\bphi=\prod_{\bx} d\phi_{\bx}.
\end{align}
% Throughout the paper, the target density is always $p\propto e^{-S}$ and the learned sampler's density is always $q$. Bold symbols denote non-scalar quantities. Plain symbols denote scalars.

\begin{figure}[!tbp]
    \centering
    \includegraphics[width=0.80\columnwidth]{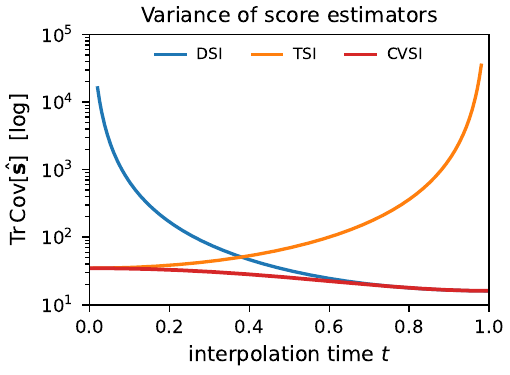}
    \caption{Score estimator variance vs.\ diffusion time $t$ ($L=4$, symmetric). While DSI and TSI diverge at either end of the time interval, CVSI variance stays low.}
    \label{fig:estimator-variance}
\end{figure}
We construct a diffusion sampler by learning to reverse a process that transforms field configurations into Gaussian noise. The reverse dynamics require the score, $\nabla \log p_t$, of the marginal distribution at each diffusion time, which we approximate with a neural network. 
% We train a score function using score matching where we regress a neural network onto the marginal score function 
%
% \begin{align}
%     &\nabla_{\bphi(t)}\log p_t(\bphi(t)) = \mathbb{E}_{p(\bphi(0)|\bphi(t))}\big[ \nabla_{\bphi(t)}\log p_t(\bphi(t)|\bphi(0)) \big] \nonumber \\
%     &=\int\nabla_{\bphi(t)}\log p_t(\bphi(t)|\bphi(0))p(\bphi(0)|\bphi(t))d\bphi(0).
%     \label{eq:marginal_score}
% \end{align} 
%
% In the following we will show two unbiased estimators of the marginal score that are typically used.
%
% We use the classical diffusion time convention throughout the paper: the diffusion time runs over $t\in[0,1]$ with
The diffusion time runs over $t\in[0,1]$ with
\begin{align}
    t=0:&\ \ \bphi(0)\sim p\propto e^{-S}
      \quad\text{(target distribution)}, \nonumber \\
    t=1:&\ \ \bphi(1)\sim p_1=\mathcal{N}(\zero,\Id)
      \quad\text{(Gaussian prior)} .
    \label{eq:time-convention}
\end{align}
Noising therefore runs \emph{forward} in $t$, from a field configuration at $t=0$ to the prior at $t=1$, and generation runs backward, from $t=1$ down to $t=0$.
The noising process is the linear Itô SDE:
\begin{align}
    d\bphi(t)=f(t)\,\bphi(t)\,dt+g(t)\,d\bw(t),
    \qquad \bphi(0)\sim p,
    \label{eq:base-sde}
\end{align}
where $f(t)$ is an affine drift, $g(t)$ is the diffusion coefficient, and $\bw$ is a Wiener process. Conditioned on its starting point $\bphi(0)$, this process has a closed-form Gaussian \emph{transition kernel}:
\begin{gather}
    p\big(\bphi(t)\,\big|\,\bphi(0)\big)
      = \mathcal{N}\big(\bphi(t);\,a(t)\,\bphi(0),\,b^2(t)\,\Id\big),
      \nonumber \\
    a(0)=1,\ b(0)=0,\ a(1)\to0,\ b(1)\to1,
    \label{eq:kernel}
\end{gather}
which we can sample directly, i.e. without simulation, during training.

Differentiating the perturbation kernel~\eqref{eq:kernel} at a fixed endpoint gives the denoising score identity (DSI):
\begin{align}
    \nabla\log p_t &= \E\big[\s_{\mathrm{DSI}}\,\big|\,\bphi(t)\big], \mbox{ where }\nonumber \\
    \s_{\mathrm{DSI}} &:= \nabla_{\bphi(t)}
      \log p\big(\bphi(t)\,\big|\,\bphi(0)\big) \nonumber \\
    &= -\frac{\bphi(t)-a(t)\bphi(0)}{b^2(t)}
     = -\frac{\beps}{b(t)}.
    \label{eq:dsi}
\end{align}
Here $\beps\sim\mathcal{N}(\zero,\Id)$ is the noise entering Eq.~\eqref{eq:kernel}. This is the standard denoising score matching (DSM) target. Forming it is cheap, since it needs only the noise $\beps$ already drawn and the configuration $\bphi(0)$, and it never touches the action. However, it has two drawbacks: First, it requires equilibrium sample from the target distribution $\bphi(0)\sim p$, which is unavailable in the data-free setting. Second, as the added noise vanishes near the physical endpoint, $t\to0$, the factor $1/b(t)$ amplifies the Gaussian noise in the regression target. Its variance therefore grows as $1/b(t)^2$, making the training signal increasingly noisy even though the underlying physical score remains finite.

% , and its $1/b(t)$ prefactor diverges at the field end $t\to0$, where $b(t)\to0$.

As \cite{de2024target} show, the marginal score can be equally rewritten as
\begin{align}
    \nabla\log p_t &= \E\big[\s_{\mathrm{TSI}}\,\big|\,\bphi(t)\big] \mbox{ where } \nonumber \\
    \s_{\mathrm{TSI}} &:= \frac{1}{a(t)}\nabla_{\bphi(0)}\log p\big(\bphi(0)\big)
      = -\frac{\nabla S\big[\bphi(0)\big]}{a(t)},
    \label{eq:tsi}
\end{align}
yielding the target score identity (TSI).
This TSM target involves the action gradient, which encodes information about the target distribution and thus enables data-free training using self-generated field configurations. Its drawback, however, is that the $1/a(t)$ pre-factor diverges at the prior end $t\to1$, where $a(t)\to0$.

DSI~\eqref{eq:dsi} and TSI~\eqref{eq:tsi} provide two unbiased estimators for the \emph{same} marginal score, which diverge at opposite ends of the schedule, as Fig.~\ref{fig:estimator-variance} shows. 
%Any convex combination is also unbiased, and CVSI~\cite{kahouli2025cvsm} uses this freedom as a control variate,
The control variate score identity (CVSI)~\cite{kahouli2025cvsm} uses the optimal combination:
\begin{align}
    \hat{\s}_{\mathrm{CVSI}}
    = \big(1-c^*(t)\big)\,\s_{\mathrm{TSI}} + c^*(t)\,\s_{\mathrm{DSI}},
    \ c^*(t)\in[0,1],
    \label{eq:cvsi}
\end{align}
\begin{figure}[!tbp]
    \centering
    \includegraphics[width=\columnwidth]{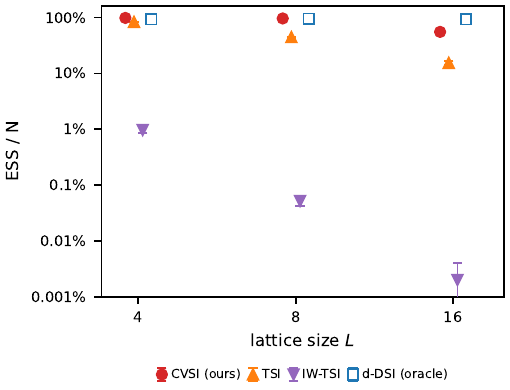}
    \caption{ESS comparison of our data-free CVSI training (red circle), TSI (orange triangle) and IW-TSI (purple inverted triangle), as well as the oracle data-driven DSI (blue square), at criticality. Error bars show 1$\sigma$ confidence intervals.}
    \label{fig:estimators-ess}
\end{figure}
%
% with $\s_{\mathrm{TSI}}$ and $\s_{\mathrm{DSI}}$ as defined above.
%(each clipped in norm before blending). 
where the coefficient $c^*(t)$ is the minimizer of the variance of the blended target. While the original formualtion $c^*$ \cite{kahouli2025cvsm} is state- and time-dependent, for a Gaussian target with per-site variance $\sigma_{\mathrm{data}}^2$ and the kernel~\eqref{eq:kernel}, this minimizer reduces to the closed form,
\begin{align}
    c^*(t) = \frac{b^2(t)}{b^2(t) + a^2(t)\,\sigma_{\mathrm{data}}^2}
           = \frac{1}{1 + \gamma^2(t)\,\sigma_{\mathrm{data}}^2},
    \label{eq:cstar}
\end{align}
where $\gamma^2(t)=a^2(t)/b^2(t)$ is the signal-to-noise ratio (Supplemental Material, Eq.~\eqref{S-eq:snr-tv}).

As shown in Fig.~\ref{fig:estimator-variance}, the CVSI variance remains finite throughout the diffusion process $t \in [0, 1]$.
At the field end $t\to 0$, 
$c^*(t)\to 0$ as 
$b(t)\to 0$, suppressing the DSI term whose $1/b(t)$ prefactor diverges, while at the prior end $t\to 1$,
$c^*(t)\to 1$ as $a(t)\to 0$, suppressing the TSI term whose $1/a(t)$ prefactor diverges.
This can also be seen from 
the explicit expression of the CVSI target~\eqref{eq:target} in the End Matter, obtained by substituting Eqs.~\eqref{eq:dsi}, ~\eqref{eq:tsi}, and \eqref{eq:cstar} into Eq.\eqref{eq:cvsi}.
For any $\sigma^2_{\mathrm{data}}>0$,
the CVSI target remains bounded throughout the entire schedule, requiring only one action-gradient computation per training point.
% CVSI finds the optimal blend, i.e. as $t\to 0$ (the field end) we have $b\to 0$ and hence $c^*\to 0$, suppressing the DSI term whose $1/b$ prefactor diverges, while as $t\to 1$ (the prior end) we have $a\to 0$ and $c^*\to 1$, suppressing the TSI term whose $1/a$ prefactor diverges.
% Substituting Eqs.~\eqref{eq:dsi}--\eqref{eq:cstar}, the blend cancels both divergences exactly for any $\sigma^2_{\mathrm{data}}>0$: the $1/a$ of the action-gradient term and the $1/b$ of the denoising term are removed, leaving a target that is bounded on the whole schedule and costs one action gradient per training point (see Eq.~\eqref{eq:target} in the End Matter for the explicit collapsed expression).

Following \cite{kahouli2025disentangling}, 
we adopt a mixed noise schedule combining variance-exploding (VE) and variance-preserving (VP) schedules.
Specifically, we match the (inverse) signal-to-noise ratio of the VE schedule, i.e. $b(t)/a(t)=\sigma(t)=\gamma^{-1}(t)$, while matching the total variance of the VP schedule, i.e. $\tau^2(t)=a^2(t)+b^2(t)=1$.
% we fix the coefficients of the kernel~\eqref{eq:kernel} by two constraints: the inverse signal-to-noise ratio of the VE-SDE, $b(t)/a(t)=\sigma(t)=\gamma^{-1}(t)$, and the total variance of the VP-SDE, $\tau^2(t)=a^2(t)+b^2(t)=1$. 
This gives
the transition kernel~\eqref{eq:kernel} with
\begin{align}
    a(t) &= \frac{1}{\sqrt{1+\sigma^2(t)}},
    \qquad
    b(t) = \frac{\sigma(t)}{\sqrt{1+\sigma^2(t)}}, \nonumber \\
    \sigma(t) &:= \sigma_{\min}\Big(\frac{\sigma_{\max}}{\sigma_{\min}}\Big)^{t}.
    \label{eq:schedule}
\end{align}

\begin{figure}[!tbp]
    \centering
    \includegraphics[width=\columnwidth]{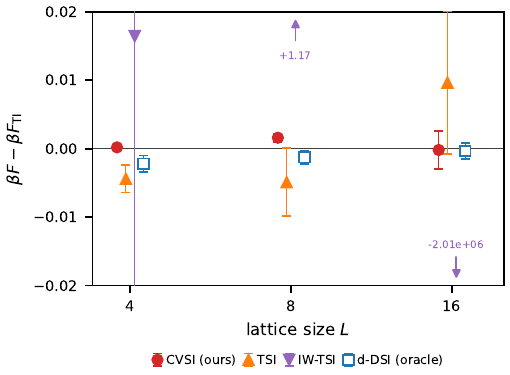}
    \caption{
    Free-energy estimation error $\beta F-\beta F_{\mathrm{TI}}$
    %of our estimator (red circle) against TSI (orange triangle), IW-TSI (purple inverted triangle), and d-DSI (oracle; blue square) 
    at criticality. 
    %Error bars show sampler $1\sigma$ uncertainty, and the grey band shows $\pm1\sigma$ uncertainty of the TI reference.
    }
    \label{fig:estimators-f}
\end{figure}

The noise level increases from $\sigma(0)=\sigma_{\min}$ at the field end to $\sigma(1)=\sigma_{\max}$ at the prior end, as required by our time convention (Supplemental Material, Eq.~\eqref{S-eq:time-convention}), so $a(0)\to1$ and $a(1)\to0$. Bounding the total variance at $\tau^2=1$ makes the prior $p_1=\mathcal{N}(\zero,\Id)$ independent of $\sigma_{\max}$ and puts $a$ and $b$ on a common scale, so that the blend coefficient of the target is set by the single scalar $\sigma^2_{\mathrm{data}}$. We use $\sigma_{\min}=10^{-2}$ throughout and treat $\sigma_{\max}$ as the one schedule hyperparameter, sweeping it below.

We train a score model $\s_{\bth}(\bphi(t),t)\approx\nabla_{\bphi(t)}\log p_t(\bphi(t))$ on pairs drawn from the kernel~\eqref{eq:kernel},
with $\bphi(0)$ drawn from the buffer. The only access we assume to the target is pointwise evaluation of $S[\bphi]$ and $\nabla S[\bphi]$.
Both score identities are available on such a pair, and we blend them as in Eq.~\eqref{eq:cvsi}. 
% Our schedule fixes $\tau^2=a^2+b^2=1$ by construction, so the blend is controlled entirely by the single scalar $\sigma^2_{\mathrm{data}}$ entering Eq.~\eqref{eq:cstar}.
In the data-free setting there is no initial dataset to read $\sigma^2_{\mathrm{data}}$ from. We therefore hold it at $\sigma^2_{\mathrm{data}}=1$ for the first $R_{\mathrm{warm}}$ rounds and thereafter re-estimate it at each buffer refresh using the model's own configurations:
\begin{align}
    \sigma^2_{\mathrm{data}} \leftarrow
      \mathrm{clip}_{[\sigma^2_-,\,\sigma^2_+]}
      \Big(\mathrm{Var}\big[\,\mathcal{B}\,\big]\Big),
    \label{eq:data-var-update}
\end{align}
where
the variance is computed over all sites across all $N_{\mathrm{gen}}$ buffer configurations. We use $R_{\mathrm{warm}}=100$ and $[\sigma^2_-,\sigma^2_+]=[0.25,10]$. We apply these safeguards during the initial rounds because the round-$0$ buffer, generated by an untrained sampler, yields variance estimates of order $10^3$--$10^5$, whereas the true variance is of order one.
% , and an unclamped estimate that large drives $c^*\to0$, i.e.\ pure TSI with no denoising term, exactly when the model is worst. 
Updating $\sigma^2_{\mathrm{data}}$ adds negligible cost, since we produce the buffer samples anyway and the variance is fast to compute.

Training alternates two loops. In the outer loop we draw $N_{\mathrm{gen}}$ configurations $\bphi(0)$ by integrating the reverse SDE from the prior at $t=1$ down to $t=0$ with the current model. These configurations replace the replay buffer $\mathcal{B}$. We also store their action gradients and re-estimate $\sigma^2_{\mathrm{data}}$. This is the only place where trajectories are simulated. In the inner loop we draw $\bphi(0)$ from $\mathcal{B}$, a time $t$ and fresh noise $\beps$, form $\bphi(t)$ in one shot with Eq.~\eqref{eq:kernel}, build the CVSI target~\eqref{eq:cvsi}, and take a gradient step. No trajectory is simulated in the inner loop and no gradient passes through an SDE solve. After $E$ steps the buffer is refreshed and the round repeats. We keep an exponential moving average of the parameters and use it for generation and evaluation. The procedure is summarised as Algorithm~\ref{alg:fixed-point} in the End Matter.

\section{Results.}
We first compare the performance of different data-free score-estimator targets and a data-driven oracle target at criticality ($\kappa = 0.2705$) on $L\in\{4,8,16\}$ lattices. The quartic coupling constant $\lambda = 0.022$ is held fixed throughout all experiments. Figures~\ref{fig:estimators-ess} and~\ref{fig:estimators-f} show ESS and the the  free-energy estimation error $\beta F-\beta F_{\mathrm{TI}}$, respectively, where the thermodynamic integration (TI) serves as the reference. 
%The comparison is between CVSI, TSI, IW-TSI, and d-DSI (oracle).
Our best choice, CVSI, ourperforms IW-TSI and TSI, which correspond to the direct application of iDEM \cite{akhoundsadegh2024idem} and adjoint matching \citep{havens2025as}, respectively, and is comparable to the  data-driven DSI (d-DSI)---the standard diffusion training on the HMC samples. 
% Our data-free CVSI estimator beats the data-free TSI estimator across all sizes, achieving $98.8\%$ at $L=4$, $96.7\%$ at $L=8$ and $55.6\%$ at $L=16$. In addition to this, it closely matches the oracle estoimator for lattice sizes \{4,8\}, i.e $98.8\%$ vs. $93.3\%$ at $L=4$, and $96.7\%$ for $95.5\%$ at $L=8$. Moreover, 
% As shown in Fig.~\ref{fig:estimators-f}, the free energy estimator using our CVSI trained model is unbiased across lattice sizes.
% , i.e. its free energy difference with the HMC reference data is $+1.99\times10^{-4}$ at $L=4$, $+1.61\times10^{-3}$ at $L=8$ and $-1.83\times10^{-4}$ at $L16$ and as shown in Tab.~\ref{tab:free-energy}, the z-score of our estimator is always between $-1$ and $+1.3$. , 
% Table~\ref{tab:free-energy} extends the main results of our estimator to all nine size/phase combinations. For $L=4$ and $L=8$, ESS exceeds $96\%$ in every phase and $32.5\%$ at $L=16$ broken phase.

We also evaluated our CVSI training in symmetric ($\kappa = 0.25$) and broken ($\kappa = 0.28$) phases.  The results are reported in 
Table~\ref{tab:free-energy}, whose right-most column gives the signed compatibility $z$-score $z=(f-f_{\mathrm{TI}})/\sqrt{\sigma_f^2+\sigma_{f_{\mathrm{TI}}}^2}$.  The estimate is considered statistically compatible with the TI reference if $|z|\lesssim 2$--$3$. 
In all cases, CVSI training yields unbiased free-energy estimates across lattice sizes and phases.

\begin{table}[!tbp]
\centering
\footnotesize
\setlength{\tabcolsep}{2.5pt}
\caption{Free energy per site $f=\beta F/|\Lambda_L|$,
its TI reference $f_{\mathrm{TI}}$,
, ESS, and the compatibility $z$-score. Errors in parentheses are $1\sigma$ on the last digits. 
Symmetric, critical, and broken pools use $N=50,000$.}
\label{tab:free-energy}
\begin{tabular}{llccrr}
\toprule
$L$ & phase & $\beta F/|\Lambda_L|$ & $f_{\mathrm{TI}}$ & ESS (\%) & $z$ \\
\midrule
4 & symmetric & $-0.687128(13)$ & $-0.687127(39)$ & 99.8 & $-0.03$ \\
 & critical & $-0.759598(22)$ & $-0.759610(54)$ & 98.8 & $+0.21$ \\
 & broken & $-0.828053(29)$ & $-0.827985(67)$ & 98.9 & $-0.93$ \\
\midrule
8 & symmetric & $-0.670360(7)$ & $-0.670380(17)$ & 99.0 & $+1.09$ \\
 & critical & $-0.721223(9)$ & $-0.721248(25)$ & 96.7 & $+0.94$ \\
 & broken & $-0.787032(11)$ & $-0.787052(35)$ & 97.4 & $+0.56$ \\
\midrule
16 & symmetric & $-0.669506(3)$ & $-0.669511(9)$ & 96.1 & $+0.46$ \\
 & critical & $-0.713112(11)$ & $-0.713111(13)$ & 55.6 & $-0.04$ \\
 & broken & $-0.778631(16)$ & $-0.778651(20)$ & 55.8 & $+0.75$ \\
\bottomrule
\end{tabular}
\end{table}

\begin{figure}[!tbp]
    \centering
    \includegraphics[width=\columnwidth]{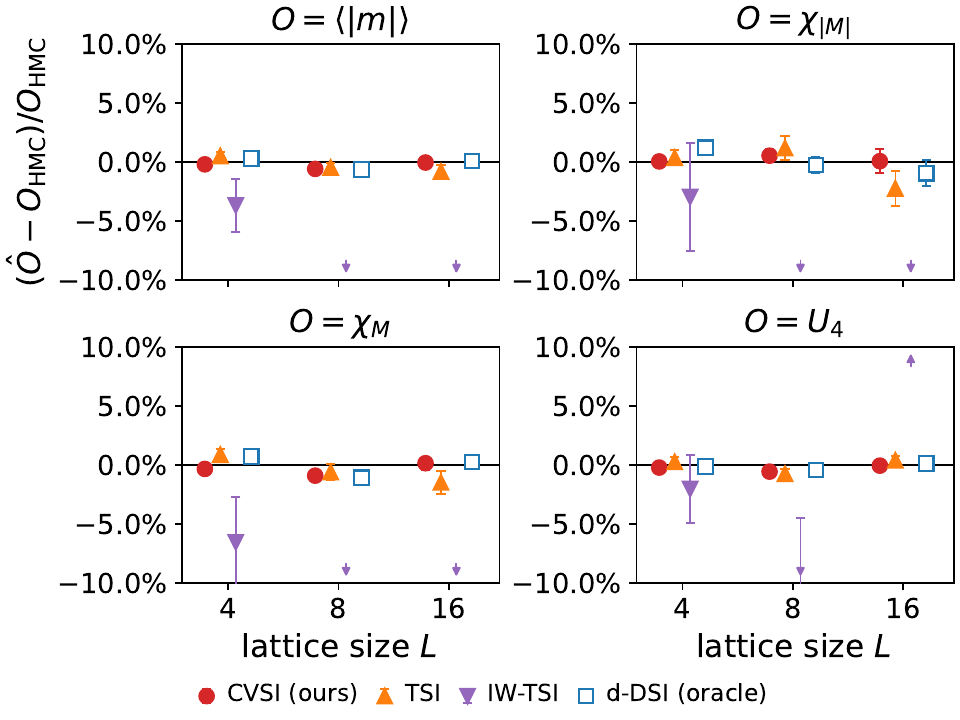}
    \caption{
    % {\color{red} Remove shadows.  Letters should be larger. Can you show results by all four methods in Fig.2?  No need to depict results out of range.  Simply mention in the caption.}
    % % Relative deviations from HMC in percent for our importance-reweighted sampler $\hat{s}$ (red circle). Error bars show sampler standard errors from joint-moment covariance propagation. Nested grey bands show HMC $1\sigma$, $2\sigma$, and $3\sigma$ errors from the $\Gamma$-method with automatic windowing ($S=2$).
    Estimation errors for the mean absolute magnetization $\langle |m| \rangle$, the susceptibility of the absolute magnetization $\chi_{|M|}$, the magnetic susceptibility $\chi_M$, and the fourth-order Binder cumulant $U_4$. 
    }
    \label{fig:observables-diff}
\end{figure}

% Finally, we examine physical observables. 
% In Figure~\ref{fig:observables-diff} shows the critical-phase residuals relative to HMC for $\langle|m|\rangle$, $\chi_{|M|}$, $\chi_M$, and $U_4$. Sampler estimates use importance weights $w_i\propto e^{-W_i}$ obtained from probability-flow ODE likelihoods, and the grey bands show the HMC $1\sigma$, $2\sigma$, and $3\sigma$ errors. Full observable results are provided in Table~\ref{S-tab:observables} of the Supplemental Material.
% \input{tables/tab_free_energy}
Figure~\ref{fig:observables-diff} shows estimation errors for other observables at criticality, using HMC sampling as the reference.  These include the mean absolute magnetization $\langle |m| \rangle$, the susceptibility of the absolute magnetization $\chi_{|M|}$, the magnetic susceptibility $\chi_M$, and the fourth-order Binder cumulant $U_4$.  For all observables and lattice sizes, the estimates are statistically compatible with the HMC reference and d-DSI oracle, suggesting that data-free training does not suffer from mode-dropping.
% the critical-phase residuals relative to HMC for . Sampler estimates use importance weights $w_i\propto e^{-W_i}$ obtained from probability-flow ODE likelihoods, and the grey bands show the HMC $1\sigma$, $2\sigma$, and $3\sigma$ errors. 
Table~\ref{S-tab:observables} in the Supplemental Material summarizes the results.

As seen in Figures~\ref{fig:estimators-ess}, importance weighting tends to degrade performance, while CVSI mitigates the main weakness of TSI: the divergence of the variance as $t \to 1$.  Our ablation study in the End Matter confirms that these trends persist across different combinations of score estimators, data-free and data-driven training, and the use or omission of importance weighting.

\begin{figure}[!tbp]
    \centering
    \includegraphics[width=\columnwidth]{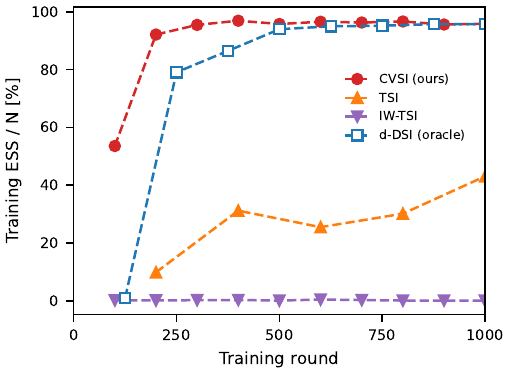}
    \caption{Training dynamics of the ESS for $N=5000$ samples.
    }
    \label{fig:iw-cvsi-train}
\end{figure}

Lastly, Fig.~\ref{fig:iw-cvsi-train} shows the ESS as a function of training steps for the critical phase at $L=8$.  We observe that IW-TSI struggles to learn the target distribution due to its high variance. CVSI accelerates the TSI training, and converges as fast as d-DSI without using HMC samples.  These results again highlight the importance of controlling target variance for stable and efficient data-free training.

\section{Conclusion.}

We have demonstrated that diffusion models can be trained for equilibrium sampling in lattice field theory without requiring reference configurations from the target distribution or a costly annealing procedure. By exploiting the target score identity (TSI), our approach enables data-free training using self-generated configurations. 
Our benchmarks in two-dimensional $\phi^4$ theory show that blending denoising score identity (DSI) with optimal balance---yielding the control variate score identity (CVSI)---significantly improves sampler performance, and importance weighting---which fixes the expectation of the target score across training steps---is harmful for stability due to its large variance.
Our suitable choice of training targets enable efficient training and sampling across the symmetric, critical, and broken phases. Remarkably, the resulting samplers achieve performance comparable to models trained on HMC reference data, demonstrating that access to target-distributed training samples is not a prerequisite for effective diffusion-based sampling. This opens a promising direction for developing generative samplers for lattice field theories where generating reference configurations is itself computationally expensive.

Several promising directions remain for future work. Since our data-free training approach is orthogonal to annealing-based methods, combining the two could further improve scalability and enable sampling on larger lattices. Its compatibility with multilevel sampling strategies also warrants investigation. Beyond scalar field theories, extending the approach to lattice gauge theories and larger lattice volumes will be important for establishing its broader applicability.

\section{Acknowledgements}

This work was supported by the German Federal Ministry of Education and Research (BMBF) under grant BIFOLD25B and by the European Union’s Horizon Europe Marie Sk\l{}odowska-Curie Doctoral Networks programme through the AQTIVATE project (grant agreement No.~101072344). 
This work is also supported with funds from the Ministry of Science, Research, and Culture of the State of Brandenburg within the Centre for Quantum Technologies and Applications (CQTA).
% This project received funding from the European Research Council (ERC) via the project ”LEEX” grant agreement 101170304 funded by the European Union. Views and opinions expressed are however those of the author(s) only and do not necessarily reflect those of the European Union or the European Research Council Executive Agency (ERCEA). Neither the European Union nor the ERCEA can be held responsible for them.
We thank Stefan K\"uhn for helpful discussions.

% \begin{acknowledgments}
% \TODO{acknowledgments}
% \end{acknowledgments}

% ══════════════════════════════════════════════════════════════
% \clearpage   % flush all body floats before the references
\bibliographystyle{apsrev4-2}
\bibliography{bibliography}

\setcounter{secnumdepth}{1}% show section numbers in the appendix
\makeatletter\let\section\revtex@section\makeatother

% --- but tighten display-math spacing in the appendix to match the main text ---
\setlength{\abovedisplayskip}{3pt plus 1pt minus 1pt}
\setlength{\belowdisplayskip}{3pt plus 1pt minus 1pt}
\setlength{\abovedisplayshortskip}{2pt plus 1pt minus 1pt}
\setlength{\belowdisplayshortskip}{2pt plus 1pt minus 1pt}

\clearpage
\phantomsection
\onecolumngrid
\begin{center}
  \Large End Matter
\end{center}
\vspace{0.8\baselineskip}

\twocolumngrid

\section{Lattice sizes, phases, network architecture, and training algorithm}

All experiments use the $D=2$ theory on periodic square lattices $\Lambda_L=\{1,\dots,L\}^2$ at fixed $\lambda=0.022$. We consider three sizes and,
through the hopping parameter $\kappa$, three phases:
\begin{align*}
    L&\in\{4,8,16\}\ \ (|\Lambda_L|=L^2\in\{16,64,256\}), \\
    \kappa&\in\{\underbrace{0.25}_{\text{sym.}},\
      \underbrace{0.2705}_{\text{crit.}},\
      \underbrace{0.28}_{\text{broken}}\}.
\end{align*}
For each of the nine $(L,\kappa)$ ensembles an HMC reference of $10^5$ configurations is used for the reference.

\begin{figure}[!b]
    \centering
    \includegraphics[width=\columnwidth]{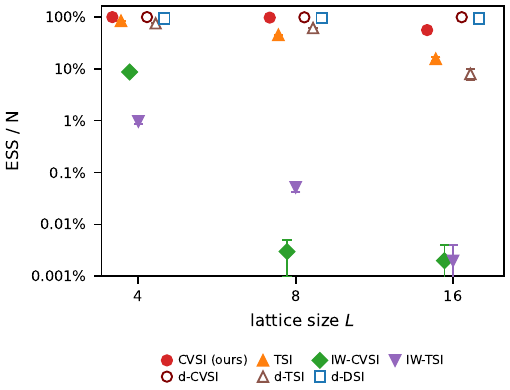}
    \caption{ESS comparison of data-free and data-driven training.}
    \label{fig:estimators-ess-endmatter}
\end{figure}

\begin{figure}[!b]
    \centering
    \includegraphics[width=\columnwidth]{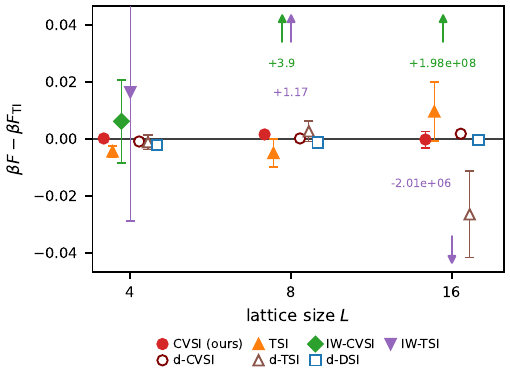}
    \caption{Free-energy estimator errors $\beta F-\beta F_{\mathrm{TI}}$.}
    \label{fig:estimators-f-endmatter}
\end{figure}

\paragraph{UNet.}
We parameterize the score with a time-conditioned convolutional UNet~\cite{ronneberger2015unet,ho2020ddpm,vega2025equivariant,Wang:2023exq,tan2026diffusion,komijani2026diffusion}, using residual blocks, attention, and skip connections. Architecture details and size-dependent settings are given in Supplemental Material, Sec.~\ref{S-sec:network-details}.
\paragraph{Circular padding.}
Spatial convolutions wrap across opposite lattice edges, implementing periodic boundary conditions.
\paragraph{$\mathbb Z_2$ symmetry.}
The score is made odd under field inversion by the model-level symmetrization
\begin{align}
 \s_{\bth}^{\mathbb Z_2}(\bphi,t)
 =\tfrac12[\s_{\bth}(\bphi,t)-\s_{\bth}(-\bphi,t)].
\end{align}

\section{The blended CVSI target and clipping}
\label{sec:target-and-clipping}
Substituting the score identities and the optimal coefficient of Eq.~\eqref{eq:cstar} gives
\begin{align}
    \hat{\s}_{\mathrm{CVSI}}\big(\bphi(0),\beps,t\big)
    = -\,\frac{a(t)\,\sigma^2_{\mathrm{data}}\,\nabla S[\bphi(0)] + b(t)\,\beps}
               {b^2(t) + a^2(t)\,\sigma^2_{\mathrm{data}}} .
    \label{eq:target}
\end{align}
For $\sigma^2_{\mathrm{data}}>0$, the blend removes the individual $1/a(t)$ and $1/b(t)$ singularities. For stability, each identity is norm-clipped before blending:
\begin{align}
    \hat{\s}_{\mathrm{CVSI}}
    &= \big(1-c^*(t)\big)\,\mathrm{clip}_{\Omega}\big(\s_{\mathrm{TSI}}\big)
     + c^*(t)\,\mathrm{clip}_{\Omega}\big(\s_{\mathrm{DSI}}\big), \nonumber \\
    \mathrm{clip}_{\Omega}(\s) &= \s\cdot\min\!\Big(\frac{\Omega}{\|\s\|},1\Big),
    \label{eq:clip}
\end{align}
We use $\Omega=70$. Clipping limits the influence of large gradients from early self-generated buffers. 

Algorithm~\ref{alg:fixed-point} summarizes the training loop.

\begin{algorithm}[tb]
\caption{Data-free training via fixed-point iteration.}
\label{alg:fixed-point}
\begin{algorithmic}
\State \textbf{Input:} $S$, $(a,b,w)$, $R$, $E$, $N_{\mathrm{gen}}$, $\eta$, $\alpha_{\mathrm{EMA}}$, $\Omega$, $R_{\mathrm{warm}}$, $[\sigma_-^2,\sigma_+^2]$
\State Initialize $\bth$, $\bar\bth\gets\bth$, $\sigma^2_{\mathrm{data}}\gets1$.
\For{$i=1,\ldots,R$}
 \State Generate $N_{\mathrm{gen}}$ fields with $\s_{\bar\bth}$ and replace $\mathcal B$.
 \If{$i>R_{\mathrm{warm}}$}
  \State Update $\sigma^2_{\mathrm{data}}$ using Eq.~\eqref{eq:data-var-update}.
 \EndIf
 \For{$e=1,\ldots,E$}
  \State Draw $\bphi(0)\sim\mathcal B$, $t\sim\mathcal U[0.01,0.99]$,
  \Statex \hspace{\dimexpr\algorithmicindent+\algorithmicindent\relax}$\beps\sim\mathcal N(\zero,\Id)$; form $\bphi(t)=a(t)\bphi(0)+b(t)\beps$.
  \State Compute $c^*(t)$ and $\hat\s$ from Eqs.~\eqref{eq:cstar}, \eqref{eq:clip}.
  \State $\bth\gets\bth-\eta\nabla_{\bth}\big[w(t)\|\s_{\bth}(\bphi(t),t)-\hat\s\|^2\big]$
  \State $\bar\bth\gets\alpha_{\mathrm{EMA}}\bar\bth+(1-\alpha_{\mathrm{EMA}})\bth$
 \EndFor
\EndFor
\State \textbf{Output:} $\s_{\bar\bth}$.
\end{algorithmic}
\end{algorithm}

% Next, we isolate the role of importance weighting in data-free training. Fig.~\ref{fig:iw-cvsi-ess} compares unweighted CVSI (ours) with its importance-weighted counterpart. IW-CVSI reaches only $8.65\%$ at $L=4$, $0.0030\%$ at $L=8$, and $0.0020\%$ in the completed 50,000-sample $L=16$ pool (effective sample size approximately one), whereas unweighted CVSI retains the efficiencies reported above. Figure~\ref{fig:iw-cvsi-train} complements this comparison with the training histories at $L=8$, including IW-TSI. Together, the comparisons motivate using the unweighted CVSI target for training.
% This choice concerns the training objective since importance reweighting remains available at evaluation to estimate physical observables from the learned sampler.

\section{Comparison to other data-driven estimators}
Figs.~\ref{fig:estimators-ess-endmatter} and \ref{fig:estimators-f-endmatter} show the comparison of our data-free training to data-driven training in terms of ESS and free-energy estimator error. We observed general trends: CVSI improves TSI and DSI, and IW substantially degrades performance.

\section{Observables and evaluation metrics}
The magnetization and its absolute value are
\begin{align}
    m[\bphi]=\frac{1}{|\Lambda_L|}\sum_{\bx\in\Lambda_L}\phi_{\bx},
    \qquad
    |m|[\bphi]=\Big| \frac{1}{|\Lambda_L|}\sum_{\bx\in\Lambda_L}\phi_{\bx} \Big|,
\end{align}
where $|\Lambda_L|=L^D$. The full distribution tests both relative mode populations and mode shapes; $\langle m\rangle\simeq0$ alone does not establish agreement with HMC.

\begin{figure}[!tbp]
    \centering
    \includegraphics[width=\columnwidth]{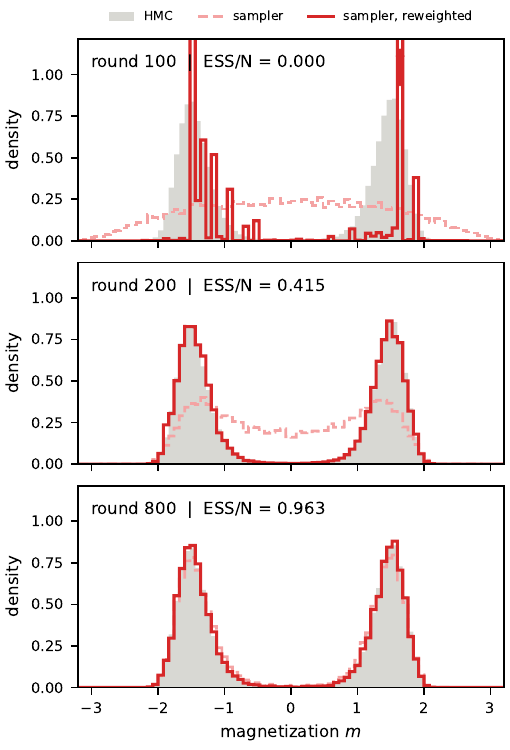}
    \caption{Magnetization at $L=8$, broken phase, over the self-learning loop. Raw samples (light red) and reweighted (red) approach the HMC reference (grey).}
    \label{fig:three-phases}
\end{figure}

Figure~\ref{fig:three-phases} shows the evolution of the raw and reweighted magnetization distributions at $L=8$ in the broken phase over the self-learning loop.

\paragraph{ESS \& Free Energy}
We compute the work for each sample, i.e.
\begin{align}
    W[\bphi]=S[\bphi]+\log q(\bphi),
\end{align}
where $q$ is the density of the learned sampler, from which the ESS and free energy are computed as
\begin{align}
    \frac{\mathrm{ESS}}{N}=\frac{\big< e^{-W} \big>^2_q}{\big< e^{-2W} \big>_q},
    \qquad
    \beta F = -\log\big< e^{-W} \big>_q .
\end{align}
The associated self-normalized importance weights are $\omega_i=\frac{e^{-W_i}}{\sum_i e^{-W_i}}$.

\end{document}